# Modification of structural and magnetic properties of $Zn_{0.96}$ $Mn_{0.04}O$ samples by $Li^{3+}$ ion irradiation

**S. K. Neogi[1], S. Chattapadhyay[2], R. Karmakar[1], Aritra Banerjee[1], S. Bandyopadhyay[1] and Alok Banerjee[3]**

[1] Department of Physics, University of Calcutta, 92 Acharya Prafulla Chandra Road, Kolkata 700009, West Bengal, India

[2]Calcutta Institute of Engineering and Management, 24/1A Chandi Ghosh Road, Kolkata 700040, India

[3] *UGC DAE Consortium for Scientific Research, Indore, Madhya Pradesh, India*

$Zn_{0.96}Mn_{0.04}O$ samples were synthesized by solid state reaction technique to explore their magnetic behavior. Structural, morphological and magnetic properties of the samples have been found to be modified by 50 MeV $Li^{+3}$ ion beam irradiation. The samples exhibit impurity phase and upon irradiation it disappears. Rietveld refinement analysis indicates that substitutional incorporation of Mn in the host lattice increases with irradiation. Grain size decreases with irradiation. Field dependent magnetization (M-H) measurement explicitly indicates ferromagnetic (FM) nature. It has been established from temperature dependent magnetization (M-T) measurement (500 Oe) and ac susceptibility ($\chi$-T) measurement that ferromagnetism in the system seems to be mainly intrinsic; though superparamagnetic Mn nanoparticles also has a minor role. The analysis of M-T data at comparatively high field (5000-Oe) provides an estimation of antiferromagnetic (AFM) exchange, which acts as a reducing agent for observed magnetic moment. The value of saturation magnetization has been increased upon irradiation and is highly correlated with dissolution of impurity phase. Actually structural property has been modified with ion irradiation and this modification may cause some definite positive change in magnetic property.

**Corresponding author*: E-mail: sbaphy@caluniv.ac.in

**INTRODUCTION:** Dilute magnetic semiconductor (DMS) is in the focus of scientific research due to their utility in spintronics device. Report of room temperature ferromagnetism in Mn doped ZnO arose the expectation of researchers in this field.[1] But this was soon followed by doubts regarding the source of the detected ferromagnetism.[2] However some conclusive reports [10,12,17] of defect induced ferromagnetism for Mn doped ZnO generates further interest among the researchers very recently. Beside there exist also some interesting reports of ferromagnetism in Mn doped ZnO samples.[3-7] In all of these reports hysteresis is observed in the M-H variation. In some cases M-T measurements were made under zero field cool (ZFC) and field cool (FC) conditions.[3-6] The distinct difference between magnetization under FC ($M_{FC}$) and magnetization under ZFC ($M_{ZFC}$) up to 300K at 50 Oe was reported.[6] Actually all of these reports[3-6] shows M-T behavior of the samples under ZFC and FC conditions with field kept in between 50 Oe to 500 Oe. We studied M-T behavior of the samples under ZFC and FC conditions at comparatively high field viz. 500 Oe and 5000 Oe. Such high field of 5000 Oe was chosen because investigation was aimed in the regime where bifurcation in $M_{FC}$ and $M_{ZFC}$ at lower field is very unusual even at low temperatures also. Generally the observed value of saturation magnetization is much smaller compared to the theoretical value of $Mn^{+2}$ state (5 $\mu_B$/Mn atom).[8] AFM coupling between neighboring Mn atoms[9-10] might be the reason for this. High field M-T measurements will enable us to explore the presence of other type of interactions, particularly AFM exchange in $Zn_{0.96}$ $Mn_{0.04}O$ samples. It has been reported from theoretical analysis that along with zinc vacancies ($V_{Zn}$), substitutional Mn at Zn sites ($Mn_{Zn}$) play active role in room temperature ferromagnetism for Mn doped ZnO samples. [21] It was shown that $Mn_{Zn}$+$V_{Zn}$ pair energetically favors FM interaction as compared with AFM interaction.[21] The experimental observation regarding the role of zinc vacancies in mediating FM interaction in Mn doped ZnO sample is reported.[7] It is totally in conformity with the theoretical standpoint.[21] So experimental observation regarding the role of $Mn_{Zn}$ in establishing ferromagnetism is quiet essential.

An important aspect in the research concerning DMS is to detect the presence of impurity phase component, if any. The analyzed sample of $Zn_{0.96}$ $Mn_{0.04}O$ consists impurity phase, $ZnMn_2O_4$. [11] The transition temperature ($T_C$) of $ZnMn_2O_4$ is around 40K. [12] So the observed ferromagnetism at 300K might not due to impurity phase contributions, rather due to original phase of the system. Ion beam irradiation was employed to dissolve impurity peaks in transition metal (TM) doped ZnO films in several cases.[13-14] But recently we have reported dissolution of impurity phase ($ZnMn_2O_4$) in case of Mn doped ZnO pellet samples also. [15] Actually in case of thin films swift heavy ions were used for irradiation but to avoid implantation in comparatively thick pellet/bulk pellet samples swift light ion of (50 MeV $Li^{+3}$) was taken for irradiation. So the aspect of dissolution of impurity phase has also been shown here. One intention of this study is to modify the structure of the sample by ion irradiation and its effect on magnetic properties. An important thing analyzed in this work: that ferromagnetism, if any, is intrinsic or impurity/ metallic cluster assisted. And finally to investigate this structural modification can enable at least some quantitative change in magnetic properties or not.

**EXPERIMENTAL DETAILS:** $Zn_{0.96}$ $Mn_{0.04}O$ pellets were synthesized by solid state reaction method. The details of sample synthesis were discussed elsewhere.[7-11] Total milling time of the samples were taken as 96 hours. One of the synthesized $Zn_{0.96}Mn_{0.04}O$ sample was irradiated with 50 MeV $Li^{3+}$ ions at room temperature using 15 UD tandem accelerator with an irradiation fluence of $10^{14}$ ions/cm$^2$. The irradiation experiment was carried out at IUAC New Delhi, India. In order to avoid the possibility of Li implantation, the irradiation were performed on samples with thickness of around 200 micron, much less than the penetration depth (220 micron), calculated from SRIM code[16] of 50 Mev $Li^{3+}$ ion beam in Mn doped ZnO samples. The un-irradiated $Zn_{0.96}Mn_{0.04}O$ [$S_1$] and irradiated $Zn_{0.96}Mn_{0.04}O$ [$S_2$] samples were investigated. X-ray diffraction (XRD), scanning electron microscope (SEM) images were employed to study structural and morphological properties of the samples. Magnetic properties were analyzed using physical property measurement system fitted with a vibrating-sample magnetometer [VSM] option. The magnetic properties were explored by M-H, M-T and $\chi$-T measurements. M-H measurements were carried at 10K and 300K. M-T measurements were made under ZFC and FC conditions at two constant magnetic fields viz. 500 Oe and 5000 Oe. A.C. susceptibility ($\chi$-T) measurements were performed at frequencies 7, 77 and 768 Hz with magnetic field kept at 3 Oe.

**RESULTS AND DISCUSSIONS:** XRD patterns of $S_1$ and $S_2$ samples have been depicted in Fig. 1 (a) and (b) respectively. Fig. 1(a) and (b) indicates polycrystalline wurtzite (hexagonal) structure of ZnO. The Fig. 1 (a) also indicate the simultaneous formation of a new phase $ZnMn_2O_4$ with emergence of a less intense peak (112) at $2\theta = 29.11^o$.

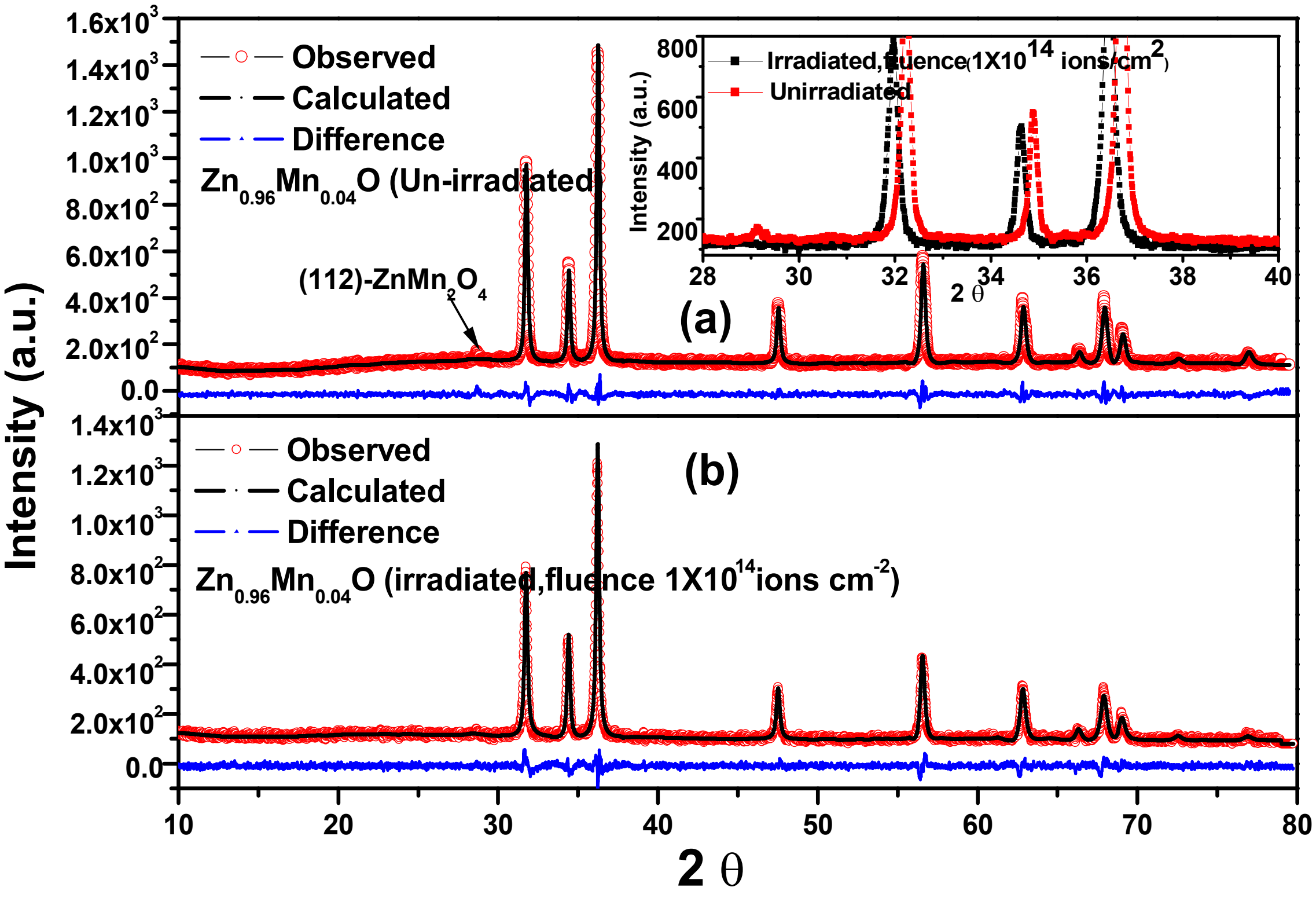

**Figure 1:** XRD pattern of $Zn_{0.96}$ $Mn_{0.04}O$ samples, (a) unirradiated and (b) irradiated with fluence of $1x10^{14}$ $Li^{3+}$ ions/$cm^2$.

Actually solubility limit is bellow 4 at% in case of Mn doped ZnO system prepared by solid state reaction method.[11-12] So it might possible that few of the Mn ions/atoms migrated out from the host ZnO matrix to form $ZnMn_2O_4$ impurity phase in $S_1$. It is noteworthy from Fig. 1(b) that the impurity peak has been dissolved within the detection limit of XRD for irradiated sample.[15] The formation and dissolution of impurity peak in $S_1$ and $S_2$ respectively is more clearly shown in the inset of Fig. 1(a) and also reported very recently.[15] The 50 MeV $Li^{3+}$ ions lose energy mainly by electronic energy loss process. Incoming energetic ions primarily deposit energy in the electronic system of the target material through excitation and ionizations of atoms. A part of this energy gets transferred to the lattice via electron– phonon interaction. As a consequence significant amount of heat is generated locally and instantaneously along the ion path, as suggested by thermal spike model, followed by rapid thermal quenching ($10^{13}$-$10^{14}$ K/s).[13] This instantaneous local heating melts the material around the path of the energetic ion that reorganized the sample by dissociating the $ZnMn_2O_4$ phase in the host ZnO matrix. There exists another interesting observation in the structural analysis; the gradual lower angle shifting of three major peaks (100), (002) and (101) for $S_2$ with respect to $S_1$ [inset of Fig. 1(a)]. The substitutional incorporation of Mn in polycrystalline ZnO was further confirmed by Reitvelt refinement analysis. This analysis was performed by fitting the measured XRD pattern of ZnO which has a wurtzite hexagonal crystal structure with space group *P*63*mc* by means of the 'Fullprof' program. The differences between the observed and fitted patterns have also been indicated in Fig. 1. The lattice parameters (a, c), unit cell volume and discrepancy factors ($\chi^2$, $R_{wp}$, $R_p$) are tabulated in Table-1. The low values of discrepancy factors indicate the credibility of the parameters.

Table1: Structural parameters estimated from Reitvelt refinement analysis

| Sample | *a* (A°) | *c*(A°) | *Volume*($A^3$) | $\chi^2$ | $R_p$ (%) | $R_{wp}$(%) |
|---|---|---|---|---|---|---|
| $Zn_{0.96}Mn_{0.04}O$-($S_1$) | 3.249 | 5.205 | 47.602 | 1.04 | 6.31 | 8.72 |
| $Zn_{0.96}Mn_{0.04}O$-($S_2$) | 3.252 | 5.213 | 47.749 | 0.452 | 4.69 | 5.97 |

It was observed that with irradiation lattice parameters and unit cell volume increases, although the proportion of Mn is same for $S_1$ and $S_2$. The radii of $Zn^{2+}$ and $Mn^{2+}$ ions are 0.60 $^0$A and 0.66 $^0$A respectively.[17] So the increase of cell volume confirm the substitutional incorporation of Mn in the host ZnO lattice. This indicates that the number of Mn ions that are outside the ZnO matrix in $S_1$ may occupy the substitutional position of Zn ion in $S_2$ due to the dissolution of $ZnMn_2O_4$ phase. So the effect of $Li^{3+}$ irradiation on structural property of Mn doped ZnO sample is quiet significant. $Mn_{Zn}$ increases with dissolution of $ZnMn_2O_4$ impurity phase for irradiated sample. So $Li^{3+}$ irradiation actually increase the solubility limit of Mn in ZnO matrix.

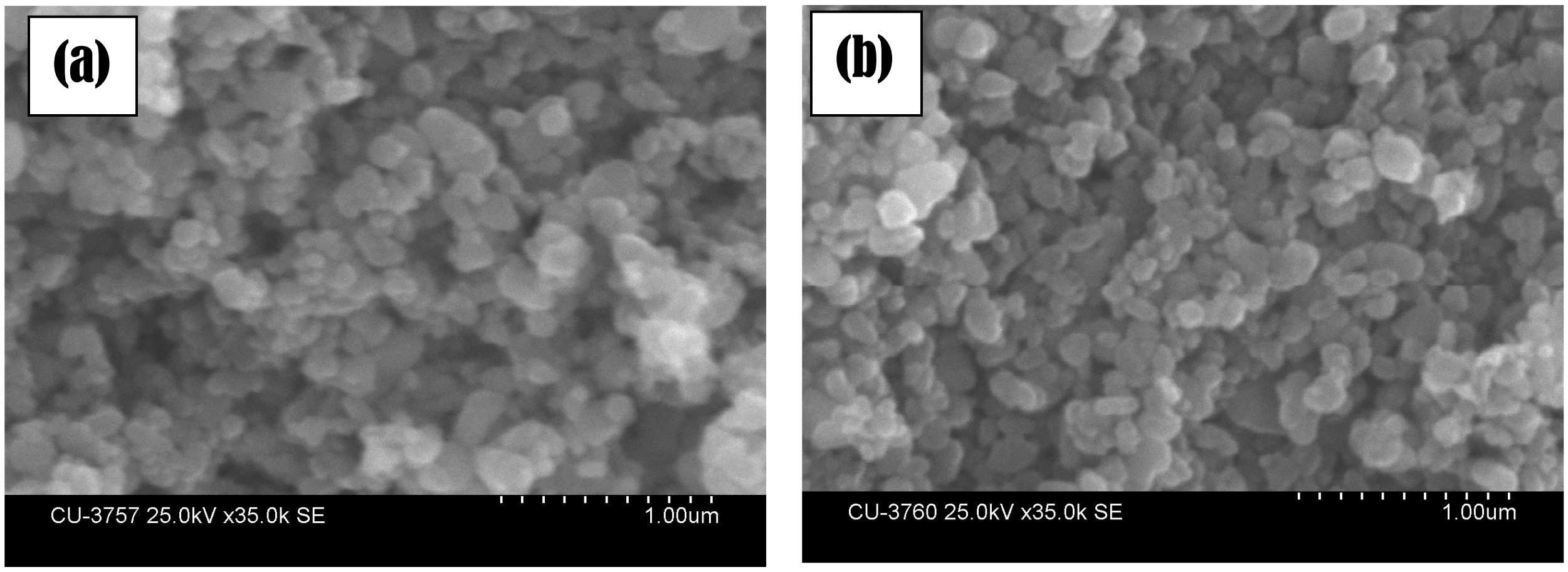

**Figure 2:** SEM micrographs of $Zn_{0.96}$ $Mn_{0.04}O$ samples, (a) unirradiated and (b) irradiated with fluence of $1x10^{14}$ $Li^{3+}$ ions/$cm^2$

SEM micrographs of $S_1$ and $S_2$ are shown in Fig. 2(a) and (b). The micrograph of $S_1$ indicates agglomeration of smaller grains to form larger ones. But with $Li^{+3}$ ion irradiation the agglomerated grains are found to some extent separated from one another. The SEM micrographs indicate distributions of particle sizes. Careful observation of SEM images indicates grain size slightly decreases with irradiation. From XRD pattern we observed decrease in peak intensity and increase in full width at half maxima (FWHM) upon irradiation for (101) peak. So the trend of variation of grain size from SEM is consistent with that of XRD.

The M-H variations of both un-irradiated and irradiated $Zn_{0.96}$ $Mn_{0.04}O$ samples at 300K & 10K have been plotted in Fig. 3(a), (b) and Fig. 4(a), (b) respectively. Inset of Fig. 3 and Fig. 4 clearly shows hysteresis behavior of $S_1$ and $S_2$ at low temperature as well as room temperature indicates the system possesses FM ordering. The phase $ZnMn_2O_4$ is not FM at room temperature so observed ferromagnetism of $S_1$ has no correlation with the impurity phase present in the sample. In the high field region M-H curves indicate that the magnetic moments are unsaturated. It might be due to coexistence of paramagnetic phase along with the ferromagnetic phase.

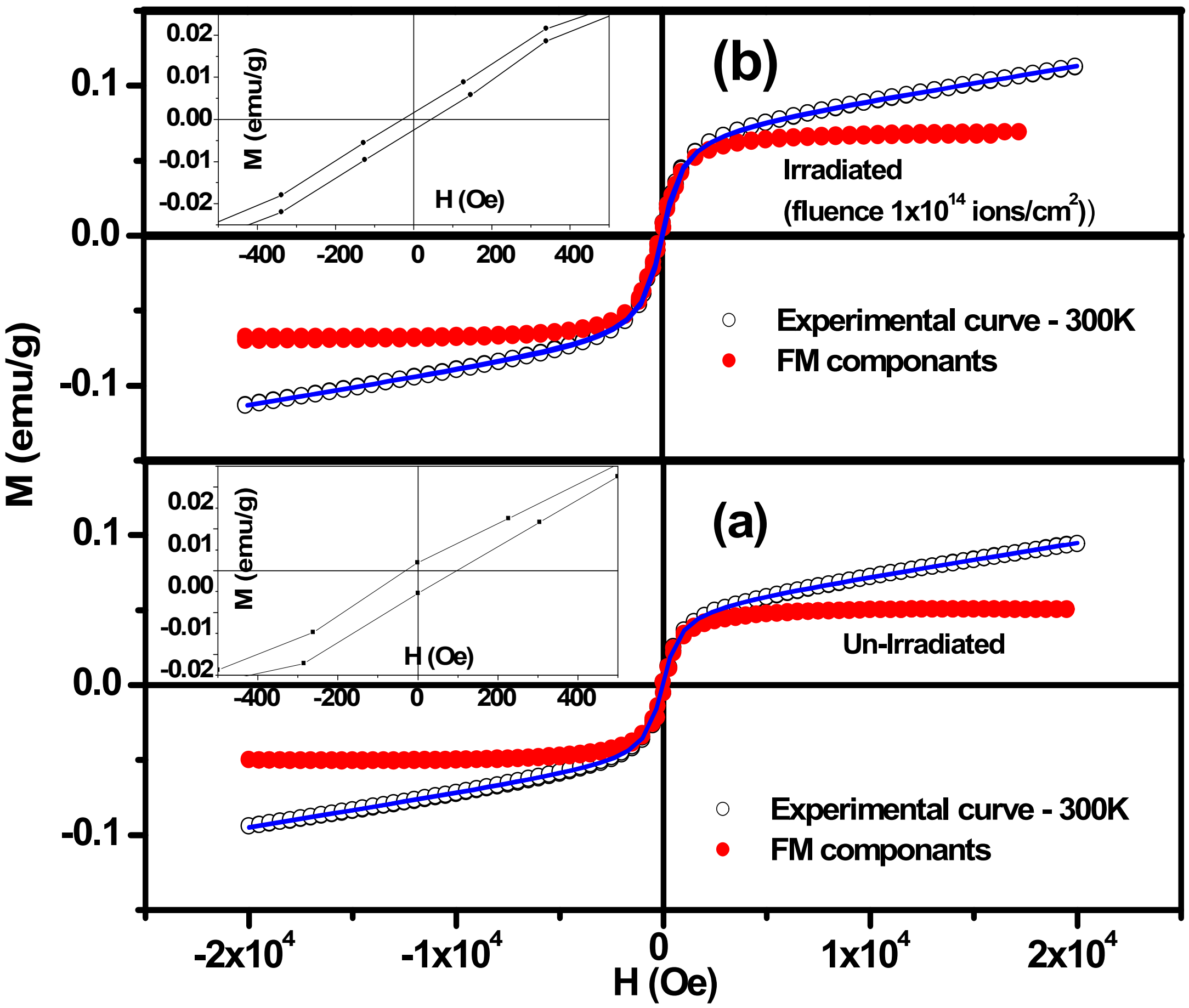

**Figure 3:** The M-H variations at 300K of $Zn_{0.96}$ $Mn_{0.04}O$ samples, (a) unirradiated and (b) irradiated with fluence of $1x10^{14}$ $Li^{3+}$ ions/$cm^2$. Inset of Fig. 3(a) & (b) Shows the expanded low field region of $S_1$ and $S_2$ at 300K. The solid line indicates the theoretical fitting of experimental data.

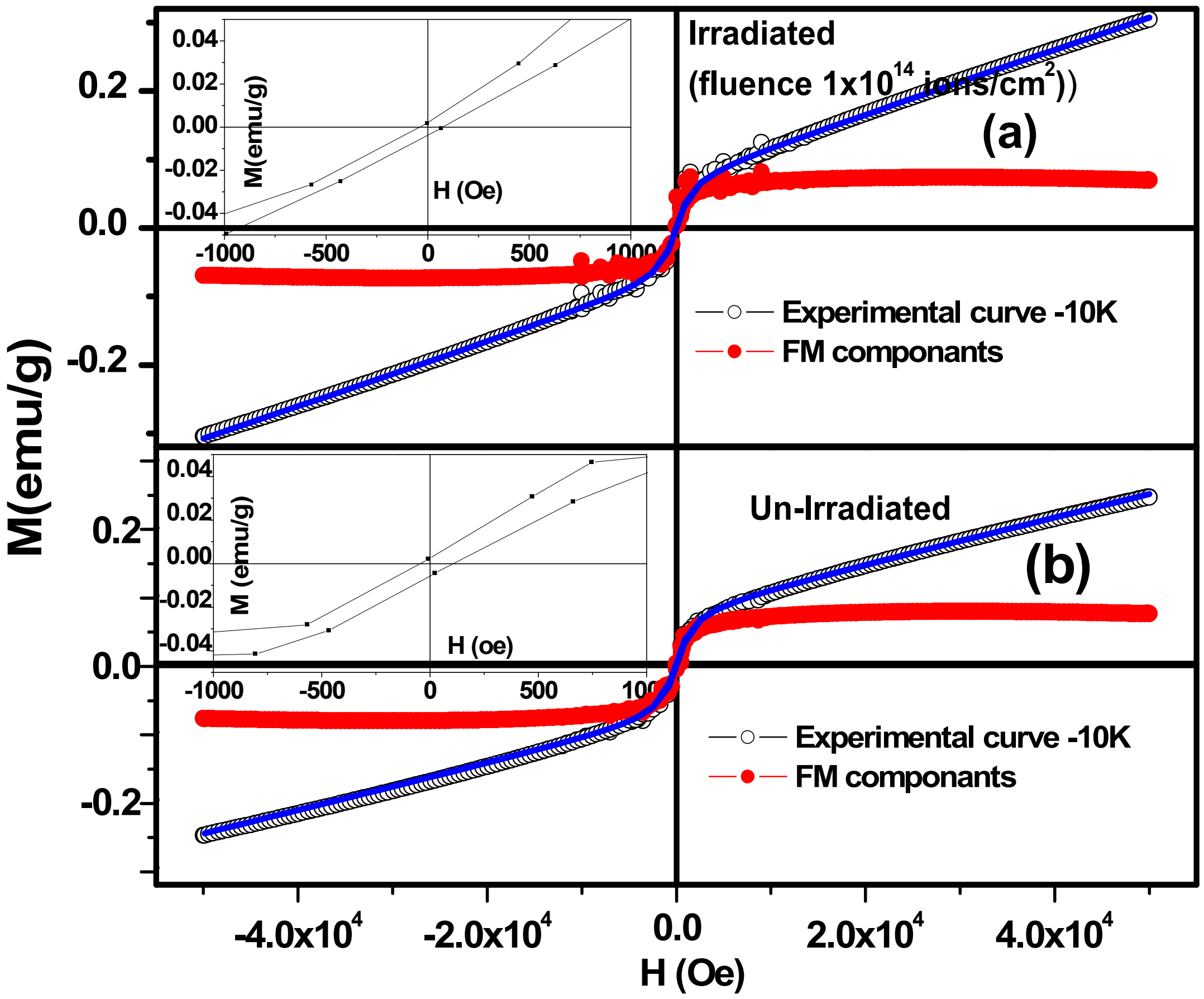

**Figure 4:** The M-H variations at 10K of $Zn_{0.96}Mn_{0.04}O$ samples, (a) unirradiated and (b) irradiated with fluence of 1x$10^{14}$ $Li^{3+}$ ions/$cm^2$. Inset of Fig. 4(a) & (b) Shows the expanded low field region of $S_1$ and $S_2$ at 10K. The solid line indicates the theoretical fitting of experimental data.

The M-H curves of both the samples did not fit well with the Brillouin-function as should be the case of purely PM or FM system. But it fits quiet well with equation (1) [18] given below. Equation (1) consists of FM as well as PM components[18] where $M_{FM}^S$, $M_{FM}^R$, $H_{ci}$ and $\chi$ are saturation magnetization, remanent magnetization the intrinsic coercivity, and PM susceptibility.

$$M(H) = (2M_{FM}^S/\pi) \tan^{-1} [(H \pm H_{ci})/H_{ci} \tan\{(\pi M_{FM}^R)/2 M_{FM}^S\}] + \chi H \ldots\ldots\ldots\ldots(1)$$

Subtracting the PM parts from the experimental data we can able to extract FM saturation of the system (as shown in Fig. 3 and Fig. 4). The values of the fitting parameters, $M_{FM}^S$ and $\chi$ of $S_1$ and $S_2$ are shown in Table-II. The estimated ratio of FM to PM contributions in $S_1$ and $S_2$ is higher in case of M-H variation at 300K. According to two-region model[19] carriers are sufficient to couple Mn ions more in FM state at 300K. At low temperature (10K) the carriers becomes localised and less no carriers push the system more towards PM state. The values of saturation magnetization ($M_{FM}^S$) are quite comparable to the reported values.[20] $S_2$ has higher value of $M_{FM}^S$ in comparison to $S_1$ indicating positive effect on magnetic property by $Li^{3+}$ ion beam. So besides introducing structural modification by dissolving the undesired impurity phase, irradiation with 50-MeV $Li^{3+}$ ions enhance desired magnetic moment values also. It might be due to more substitutional incorporation of Mn at zinc site ($Mn_{Zn}$) with irradiation as evident from increase in cell volume (estimated from Reitvelt refinement analysis of XRD spectra). This observation strongly support the theoretical viewpoint that formation of $Mn_{Zn}$ energetically favors FM exchange.[21] Actually with dissolution of impurity phase excess Mn atoms migrate into the host ZnO matrix by substitutionally replacing Zn atoms and hence the value of $M_{FM}^S$ increases. So structural modification arising out of dissolution of impurity phase might be the cause for substantial up-gradation of magnetic properties with irradiation.

Table II: Magnetic parameters estimated from M-H and M-T (at 5000 Oe) measurements

| Sample | $M_{FM}^{S}$($\mu_B$/Mn) -10K | $M_{FM}^{S}$($\mu_B$/Mn) - 300K | χ(emu/g-Oe) 10K | χ(emu/g-Oe) 300K | *C*(*x*) | *Θ*(*x*) | $J_1/k_B$(°K) |
|---|---|---|---|---|---|---|---|
| $Zn_{0.96}Mn_{0.04}O$-($S_1$) | 0.029 | 0.019 | $3.40 \times 10^{-6}$ | $2.15 \times 10^{-6}$ | 0.0129 | 815.5 | 291 |
| $Zn_{0.96}Mn_{0.04}O$ ($S_2$) | 0.027 | 0.026 | $4.67 \times 10^{-6}$ | $2.17 \times 10^{-6}$ | 0.0135 | 810.6 | 298 |

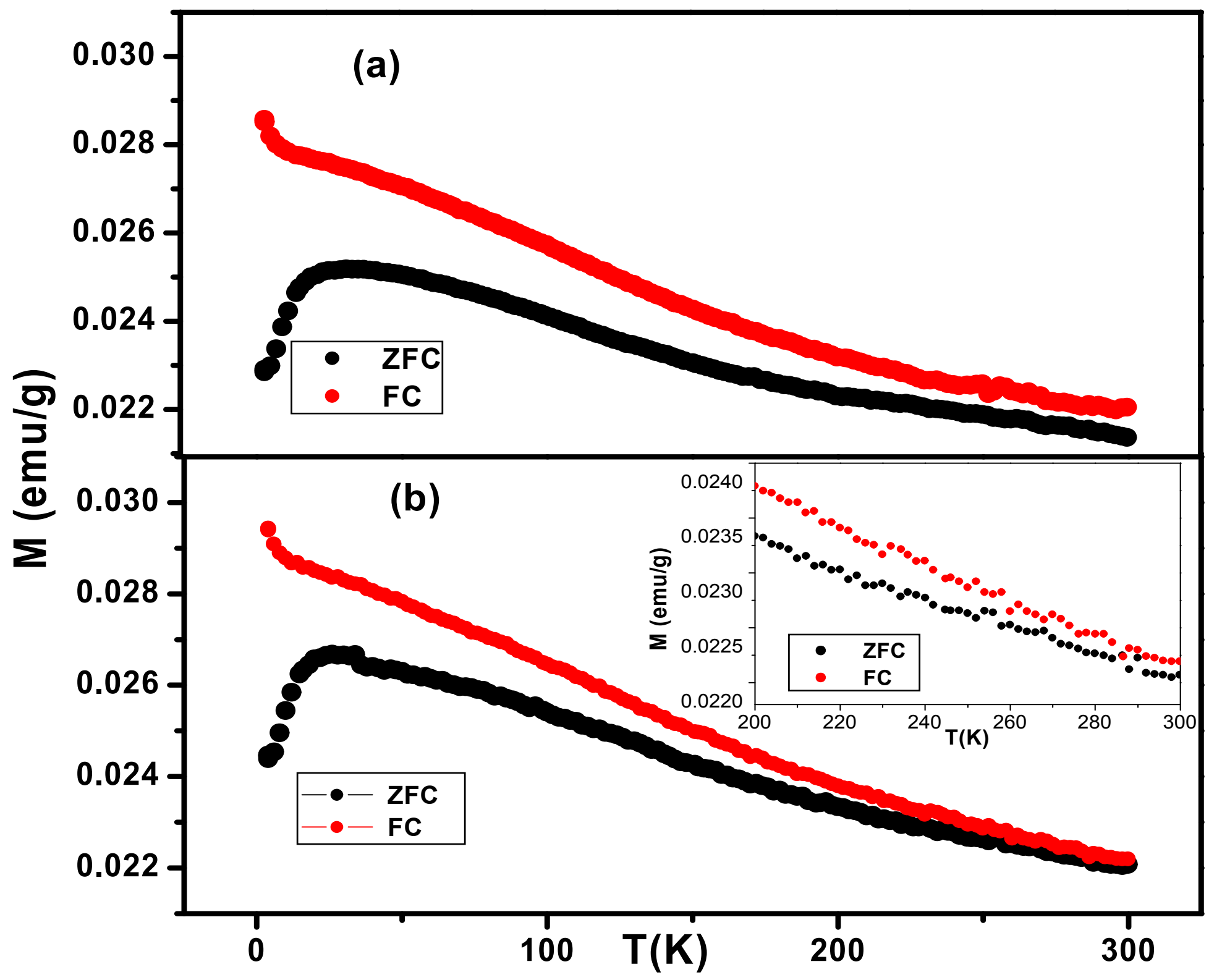

**Figure 5**: The ZFC FC data taken at 500Oe field of $Zn_{0.96}$ $Mn_{0.04}O$ samples (a) unirradiated and (b) irradiated with fluence of $1 \times 10^{14}$ $Li^{3+}$ ions/cm$^2$.Inset of Fig. 5(b) shows enlarged portion of the ZFC FC moments of irradiated $Mn_{0.04}Zn_{0.96}O$.

M-T variation of $S_1$ and $S_2$ at 500 Oe under ZFC and FC conditions has been shown in Fig. 5 (a) and (b). $M_{ZFC}$ curves indicate gradual increase in low temperatures, reach maximum around spin freezing or blocking temperature ($T_B$) and then gradually decreases with rise in temperature. But $M_{FC}$ curves continuously decreases with rise in temperature. This type of nature is observed in superparamagnetic/spin-glass system. [22] But a distinct difference exists between $M_{FC}$ and $M_{ZFC}$ curves up to 300K in both cases. The clear difference between $M_{FC}$ and $M_{ZFC}$ curves well above the $T_B$ and the observed hysteresis in the M-H curve at 300K, distinguishes this system from conventional superparamagnetic/spin-glass system, where $M_{FC}$ curve merges together with the $M_{ZFC}$ curve just at or above $T_B$. The observed broad peak in $M_{ZFC}$ curve could be due to size distributions of the Mn nano particles.[22-24] So presence of minute amount of superparamagnetic Mn nanoparticles cannot be ruled out in these samples. But at the same time it is noteworthy that room temperature FM ordering manifested by these samples is due to the original phase of the system. The formation of Mn:ZnO based dilute magnetic semiconductor is primarily responsible for FM ordering[25] otherwise $M_{FC}$ and $M_{ZFC}$ curves merges at or above $T_B$. So FM ordering seems to be intrinsic here. The difference between $M_{FC}$ and $M_{ZFC}$ curves at 300K shows that $T_C$ is above the room temperature for both $S_1$ and $S_2$. But more separation between $M_{FC}$ and $M_{ZFC}$ curves at 300K pointing towards higher $T_C$ value in case of $S_1$ in comparison to $S_2$. Careful inspection indicates that upon irradiation only $M_{ZFC}$ values are increased whereas $M_{FC}$ values remain same to that of un-irradiated sample. Actually clear separation between $M_{FC}$ and $M_{ZFC}$ is desired for intrinsic ferromagnetism and that is observed for both $S_1$ and $S_2$. More separation between $M_{FC}$ and $M_{ZFC}$ at 300K in $S_1$ might not be indicated as stronger FM nature. Rather it indicates stronger inter particle interaction in $S_1$[27]. Besides it has been reported [7] that FM ordering increases with increasing concentration of defects specifically Zn vacancy. Structural modification induced by $Li^{3+}$ ion beam irradiation may change the nature of the defect states in the sample. Already irradiated sample exhibits better Mn incorporation in the host ZnO matrix as evident from the structural analysis. So a complete reorientation of defect states has been taken place for irradiated sample. It might be another cause of narrowing in separation of $M_{FC}$ and $M_{ZFC}$ curves at 300K in case of irradiated specimen.

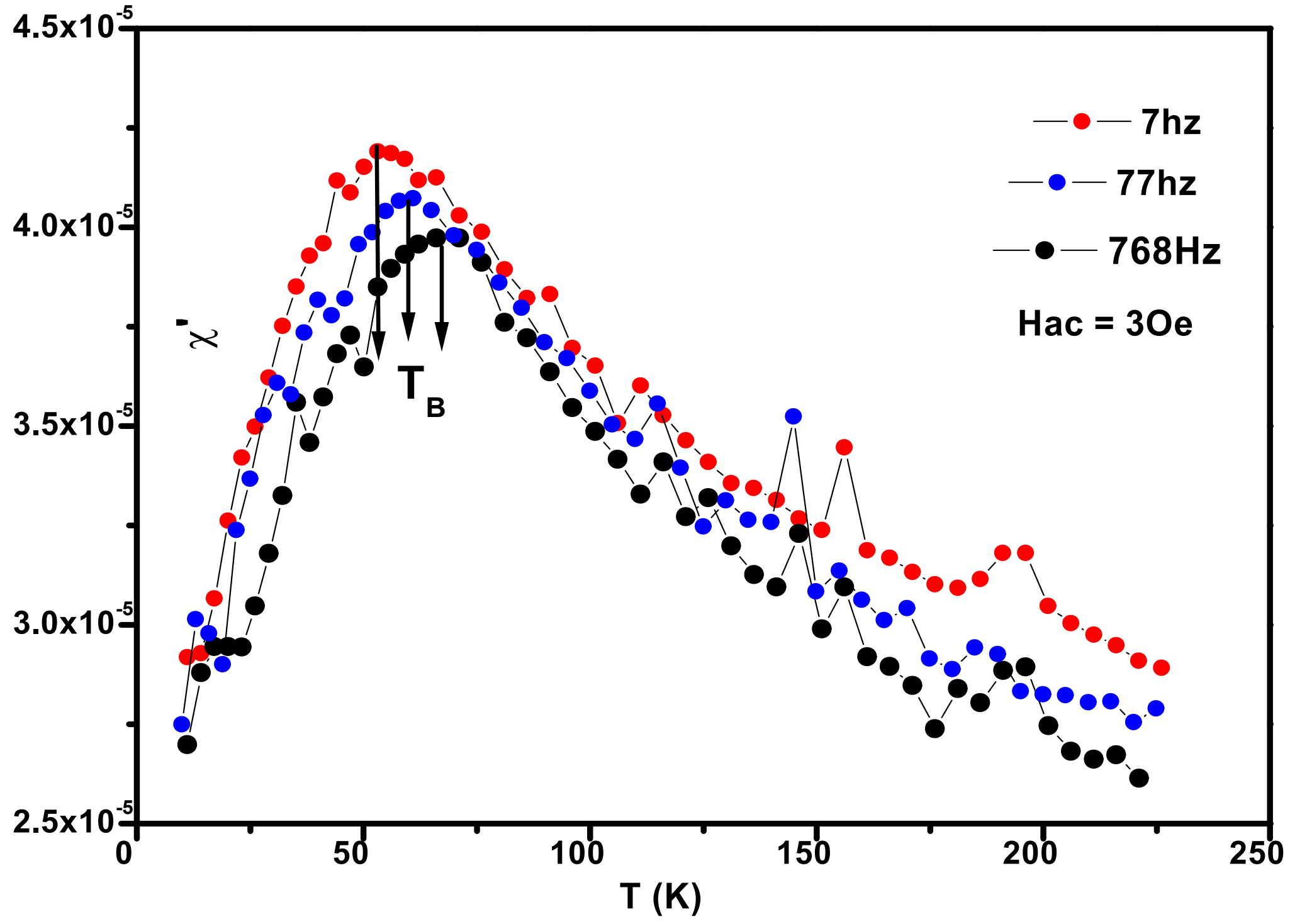

**Figure 6:** The temperature dependence of real part of ac susceptibility (χ) at 3Oe with frequency 7, 77, and 768 Hz as parameter (from top to bottom) of unirradiated $Zn_{0.96}$ $Mn_{0.04}O$ sample.

Fig. 6 shows the real part of magnetic susceptibility ($\chi^1$) of unirradiated $Zn_{0.96}$ $Mn_{0.04}O$ has been plotted against temperature with frequency as parameter. It shows magnetic susceptibility attains a broad peak and then slowly decreases with increasing temperature. The peak gradually shifts towards higher temperature with increasing frequency. It indicates slowing down of the magnetization switching occurred at blocking temperature, which is typical of superparamagnetic system.[3] Further, the substantial shift of the peak of $\chi_1$ towards higher temperature with increasing frequency and slow decrease after the maxima eliminates the possibility of the system to be in spin-glass state rather to be in superparamagnetic state. But in case of typical superparamagnetic system curves of $\chi_1(T)$ at different frequencies merge with one another in the entire temperature range except around maxima (blocking temperature).[24,26] Though small but still distinct separation exists here between all three curves of $\chi_1(T)$ in the entire temperature range. It indicates presence of superparamagnetic Mn nanoparticles in the system. But primary reason of ferromagnetism is again attributed to be as Mn:ZnO based DMS which is consistent with the observation of M-T measurement at 500 Oe.

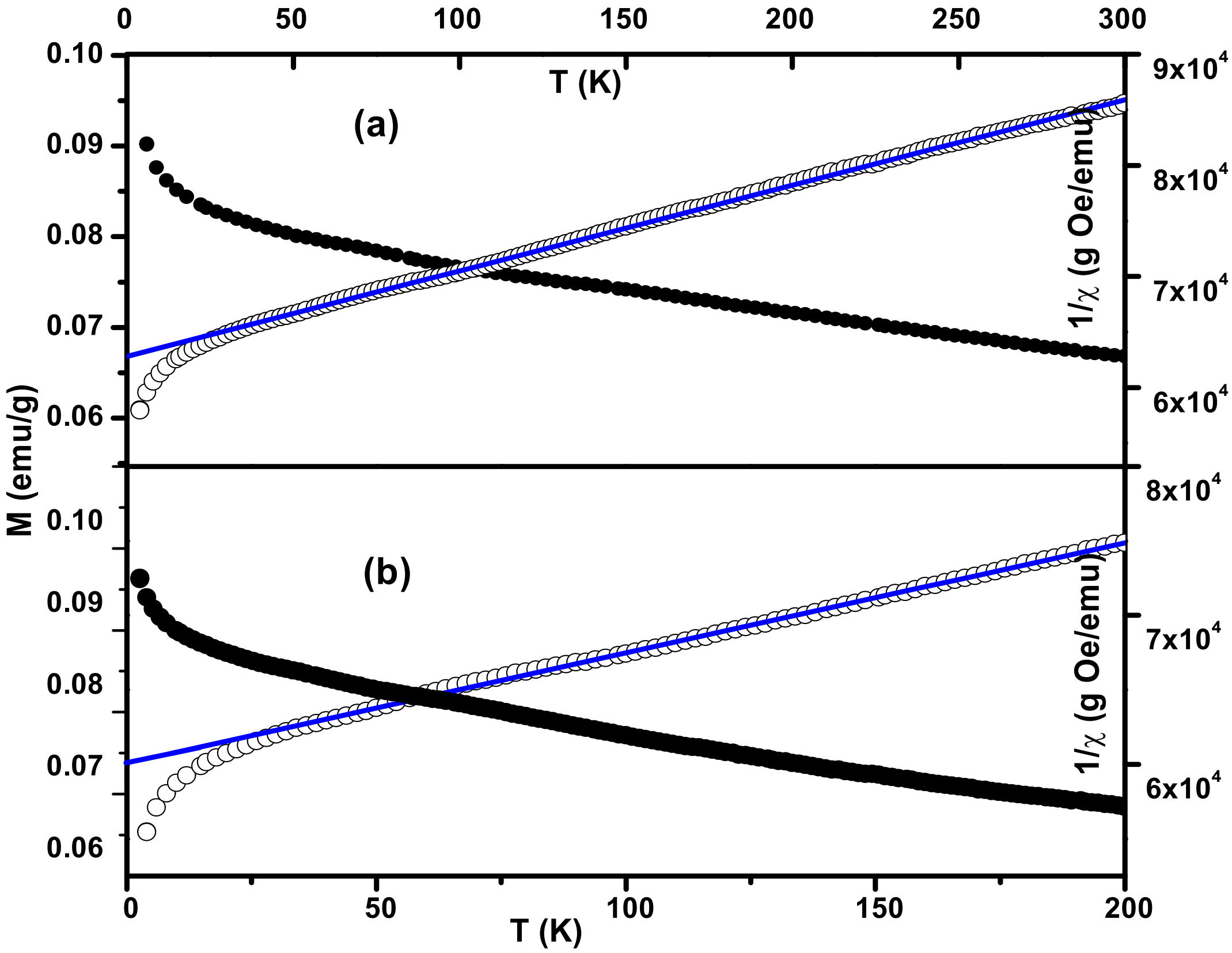

**Figure 7:** Temperature dependent magnetization ($M_{ZFC}$) and inverse susceptibility against temperature [$\chi^{-1}(T)$] variations of of $Zn_{0.96}$ $Mn_{0.04}O$ samples, (a) unirradiated and (b) irradiated with fluence of $1x10^{14}$ $Li^{3+}$ ions/cm$^2$.

M-T measurements of $S_1$ and $S_2$ have also been performed at field strength of 5000Oe under ZFC and FC conditions. Any separation between $M_{ZFC}$ and $M_{FC}$ values in the entire temperature range has not been found. This trend is not quiet astonishing. It can be understood by realizing that at a given temperature, $M_{FC}$ - $M_{ZFC}$ = 0 if the measuring field is higher than the maximum field at which hysterisis is observed in M(H) loops. The variation of inverse of magnetic susceptibility ($1/\chi$) (estimated from $M_{ZFC}$ values at 5000 Oe) with temperature has been shown in Fig. 7 (a) and (b) for respective samples.

The $1/\chi$-T curve exhibits typical antiferromagnetic behavior. The straight line represents least square fit of Curie-Weiss law to the experimental data in the temperature range 80 to 300K. The Curie-Weiss law is expressed as.[28-30]

$$M/H = C(x)/\{T- \Theta(x)\} \quad (2)$$

Curie constant $C(x)$ is given by $C(x) = xC_o = x(g\mu_B)^2S(S+1)n/3k_B$ (3)

the Curie–Weiss temperature $\Theta(x)$ expressed as $\Theta(x) = x\Theta_0 = 2xS(S+1)zJ_1/3k_B$ (4)

where n is the no of Mn atoms/ volums, g=2.0, $\mu_B$=9.27x10$^{-21}$ergs/Oe, and $k_B$=1.38x10$^{-16}$ ergs/K, and x is the concentration of Mn atoms, S the value of the spin for the $Mn^{2+}$ ions in ZnO, $J_1$ is the effective exchange integral between nearest neighbor $Mn^{2+}$ ions and z is the no of nearest neighbor cations (= 12 for wurzite structure). The fitting of $1/\chi$ against T is quiet good. The values of $C(x)$, $\Theta(x)$ have been estimated from slope and intercept of equation (2). The value of $J_1/k_B$ has been estimated from equation (4) using the value S=5/2. The values of $C(x)$, $\Theta(x)$, and $J_1/k_B$ are shown in Table 2. The large values of $\Theta(x)$ suggest presence of strong AFM coupling.[29-30] Actually in competition between two spontaneous interactions i.e. FM and AFM exchange, stronger FM interaction dominates and that is on an average reflected in M-H measurement. There are reports[5-7] that indicate observed values of saturation magnetization is unexpectedly low as compared to theoretically predicted value ($5\mu_B$/Mn atom).[8] Primarily the reason attributed for such low values of magnetization is the presence of AFM coupling between neighboring Mn atoms.[9-10] Here we quantitatively estimate the AFM coupling by analyzing the high field (5000 Oe) MT data of $Zn_{0.96}$ $Mn_{0.04}O$ (both $S_1$ and $S_2$) samples.

**CONCLUSION:** To summarize, both of the samples are FM in nature. The un-irradiated sample has $ZnMn_2O_4$ impurity phase. The $ZnMn_2O_4$ phase has been dissolved upon ion irradiation. Rietveld refinement analysis of structural data indicates greater incorporation of Mn in ZnO matrix for irradiated sample. The observation of hysteresis loop at 10K and even at 300K is the prime feature in M-H measurement. The coupled PM component has been decoupled from FM counterpart by mathematical modeling. Though presence of minute amount of superparamagnetic Mn nanoparticles cannot be ruled out but the main factor behind this FM ordering has been identified as Mn:ZnO based DMS. The less separation between $M_{FC}$ and $M_{ZFC}$ curves at 300K in $S_2$ is due to weaker inter particle interaction and reorientation of defect states upon irradiation. The presence of strong AFM coupling has been estimated. This AFM coupling reduces the magnetic moment of the samples from the theoretically predicted value. The comparatively high value of saturation magnetization was observed for $S_2$. Actually dissolution of impurity phase with ion irradiation excess Mn atoms migrates towards the position of substitutional zinc site and might be the cause of this increase. It indicates structural modification arising from $Li^{3+}$ ion irradiation imparted some positive effect on the overall magnetic properties of the sample.

*This work is financially supported by DST-Government of India, and IUAC-New Delhi, vide project nos: SR/FTP/PS-31/2006 and UFUP-43308 and CRNN-University of Calcutta. The author S.K.N. is thankful to UGC for providing his fellowship. We acknowledge Department of Chemical Technology-University of Calcutta, UGC-DAE CSR-Indore and IUAC-New Delhi to use their Scanning electron microscope (SEM), magnetic measurement and pelletron facilities respectively.*

**References**

[1] P. Sharma, A. Gupta, K.V. Rao, F. J. Owens, R. Sharma, R. Ahuja, J. M. O. Guillen, B. Johansson, and G. A. Gehring, Nat. Mater. **2**, 673 (2003).

[2] D. P. Norton, M. E. Overberg, S. J. Pearton, K. Pruessner, J. D. Budai, L. A. Boatner, M. F. Chisholm, J. S. Lee, Z. G. Khim, Y. D. Park, and R. G. Wilson, Appl. Phys. Lett. **83**, 5488 (2003).

[3] T. Meron, and G. Markovich, J. Phys. Chem. B **109,** 20232 (2005).

[4] J. Zhang, R. Skomski and D. J. Sellmyer, J. Appl. Phys. **97**, 10D303 (2005).

[5] O. D. Jayakumar, H. G. Salunke, R. M. Kadam, M. Mohapatra, G. Yaswant and S. K. Kulshreshtha, Nanotechnology, **17,** 1278 (2006).

[6] O. D. Jayakumar, I. K. Gopalakrishnan, C. Sudakar and S. K. Kulshreshtha, J. Cryst. Growth, **294,** 432 (2006).

[7] S. Chattopadhyay, S. K. Neogi, A. Sarkar, M. D. Mukadam, S. M. Yusuf, A. Banerjee and S. Bandyopadhyay, J. Magn. Mag. Mats. **323**, 363 (2011).

[8] T. Dietl, H. Ohno, F. Matsukura, J. Cibert, D. Ferrand, Science **287,** 1019 (2000).

[9] S. Ramachandran, J. Narayan, J. T. Prater, Appl. Phys. Lett. **88,** 242503 (2006).

[10] Y. W. Heo, M. P. Ivill, K. Ip, D. P. Norton, S. J. Pearton, J. G. Kelly, R. Rairigh, A. F. Hebard, T. Steiner, Appl. Phys. Lett. **84,** 2292 (2004).

[11] S. Chattopadhyay, S. Dutta, A. Banerjee, D. Jana, S. Bandyopadhyay, S. Chattopadhyay, and A. Sarkar, Physica B **404**,1509 (2009).

[12] J. Blasco, F. Bartolome, L. M. Garcıa and J. Garcıa J. Mater. Chem. **16**, 2282, (2006).

[13] B. Angadi, Y.S.Jung, W.K.Choi, R. Kumar, K. Jeong, S.W.Shin, J.H.Lee, J.H.Song, M.W.Khan and J.P.Srivastava, Appl. Phys. Lett. **88,** 142502 (2006).

[14] R. Kumar, A. P. Singh, P. Thakur, K. H. Chae, W. K. Choi, B. Angadi, S. D. Kaushik and S. Patnaik, J. Phys. D: Appl. Phys. **41**,155002 (2008).

[15] S K Neogi, S. Chattopadhya, Aritra Banerjee,S. Bandyopadhyay, A. Sarkar and Ravi Kumar. J. Phys.: Condens. Matter **23** 205801 (2011).

[16] Ziegler, F.J.; Biersack, J. P.; Littmerk, U. Stopping Power and Ranges of Ion in Matter (SRIM), Pergamon Press, New York, 1985; Ziegler, J. F.; Ziegler, M. D.; Biersack, J. P. Computer code SRIM available from http://www.SRIM.org

[17] D. Rubi, J. Fortcubetra, A. Calleje, L. Aragones, X. G. Capdevile M. Segarra, Phys. Rev. B. **75**, 155322 (2007).

[18] S. Duhalde, M.F. Vignolo, and F. Golmar, C. Chiliotte, C. E. Rodríguez Torres, L. A. Errico, A. F. Cabrera, M. Rentería, and F. H. Sánchez and M. Weissmann, Phys. Rev. B **72**,161313 (2005).

[19] M.J. Calderon and S. Das Sarma, Ann. Phys. **322**, 2618 (2007).

[20] B. B. Straumal, A. A. Mazilkin, S. G. Protasova, A. A. Myatiev, P. B. Straumal, G. Schütz, P. A. van Aken, E. Goering, and B. Baretzky, Phys. Rev. B **79**, 205206 (2009).

[21] W. Yan, Z. Sun, Q. Liu, Z. Li, Z. Pan, D. Wang, Y. Zhou, X. Zhang, Appl. Phys. Lett. **91,** 062113(2007).

[22] M. Respaud, J. M. Broto, H. Rakoto, A. R. Fert, L. Thomas and B. Barbara M. Verelst, E. Snoeck, P. Lecante, A. Mosset, J. Osuna, T. Ould Ely, C. Amiens, and B. Chaudret, Phys. Rev. B **57**, 2925 (1998).

[23] J. H. Park, M.G. Kim, H. M. Jang, S. Ryu, and Y. M. Kim, Appl. Phys. Lett **84**, 1338 (2004),

[24] S. Zhou, K. Potzger, J. von Borany, R. Grotzschel, W. Skorupa, M. Helm, and J. Fassbender, Phys. Rev. B, **77**, 035209 (2008)

[25] M. Opel, K.W. Nielsen, S. Bauer, S.T.B. Goennenwein, J.C. Cezar, D. Schmeisser, J. Simon, W. Mader, and R. Gross, Eur. Phys. J. B **63**, 437 (2008).

[26] S. Zhou, K. Potzger, G. Zhang, F. Eichhorn W. Skorupa, M. Helm, and J. Fassbender. J. Appl. Phys, **100**, 114304 (2006).

[27] S. Bhattacharyya and A. Gedanken, J. Phys. Chem. C. **112**, 4517 (2008).

[28] J. Spalek, A. Lwicki, Z. Tarnawski, J.K. Furdyna,R.R.Galazka and Z. Obuszko Phys. Rev. B **33,** 3407 (1986).

[29] S. Banerjee, K. Rajendran, N. Gayathri, M. Sardar, S. Senthilkumar, and V. Segodan, J. Appl. Phys. **104**, 043913 (2008).

[30] J. Luo, J. K. Liang, Q. L. Liu, F. S. Liu, Y. Zhang, B. J. Sun, and G. H. Rao, J. Appl. Phys**. 97,** 086106 (2005).